\documentclass[confrence]{IEEEtran}
\IEEEoverridecommandlockouts
\usepackage{cite}
\usepackage{amsmath,amssymb,amsfonts}
\usepackage{algorithm}
\usepackage{algorithmic}
\usepackage{setspace}
\usepackage{graphicx}
\usepackage{textcomp}
\usepackage{xcolor}
\usepackage{stfloats}
\usepackage{graphicx}
\usepackage{color}
\usepackage{multirow}
\usepackage{array}
\usepackage{bm}
\usepackage{subfig}
\usepackage{booktabs}
\usepackage{caption}
\usepackage{lettrine}

\usepackage{geometry}
\usepackage{authblk}

\def\BibTeX{{\rm B\kern-.05em{\sc i\kern-.025em b}\kern-.08em
		T\kern-.1667em\lower.7ex\hbox{E}\kern-.125emX}}
\begin{document}
	
	\title{A Graph-Based Collision Resolution Scheme for Asynchronous Unsourced Random Access\\}
	\author{\IEEEauthorblockN{Tianya Li, Yongpeng Wu, Wenjun Zhang, Xiang-Gen Xia, and Chengshan Xiao}
		\vspace{-0.4cm}
		\thanks{{The work of Y. Wu is supported in part by the National Key R\&D Program of China under Grant 2018YFB1801102, the Fundamental Research Funds for the Central Universities, National Science Foundation (NSFC) under Grant 62122052 and 62071289, 111 project BP0719010, and STCSM 22DZ2229005.
				
				T. Li, Y. Wu (corresponding author) and W. Zhang are with the Department of
				Electronic Engineering at Shanghai Jiao Tong University, Shanghai, China.
				Emails: \{tianya, yongpeng.wu, zhangwenjun\}@sjtu.edu.cn.
							
				 X.-G. Xia is with the Department of Electrical, and Computer Engineering, University of Delaware, Newark, DE 19716 USA. E-mail: xianggen@udel.edu.
				
				C. Xiao is with the Department of Electrical, and Computer Engineering, Lehigh University, Bethlehem, PA 18015 USA. E-mail: xiaoc@lehigh.edu.
		}}
	}

\maketitle
\thispagestyle{empty}

\begin{abstract}

\par This paper investigates the multiple-input-multiple-output (MIMO) massive unsourced random access in an asynchronous orthogonal frequency division multiplexing (OFDM) system, with both timing and frequency offsets (TFO) and non-negligible user collisions. The proposed coding framework splits the data into two parts encoded by sparse regression code (SPARC) and low-density parity check (LDPC) code. Multistage orthogonal pilots are transmitted in the first part to reduce collision density. Unlike existing schemes requiring a  quantization codebook with a large size for estimating TFO, we establish a \textit{graph-based channel reconstruction and collision resolution (GB-CR$^2$)} algorithm to iteratively reconstruct channels, resolve collisions, and compensate for TFO rotations on the formulated graph jointly among multiple stages. We further propose to leverage the geometric characteristics of signal constellations to correct TFO estimations. Exhaustive simulations demonstrate remarkable performance superiority in channel estimation and data recovery with substantial complexity reduction compared to state-of-the-art schemes.
\end{abstract}


\section{Introduction}

\par \lettrine[lines=2]{A}{iming} to provide massive connectivity for the burgeoning communication services with short packets and sporadic traffic, massive machine-type communications (mMTC) has been a pivotal application scenario in the fifth-generation (5G) wireless communication \cite{wu2020wcm}. As a prospective protocol, grant-free random access (GF-RA) has recently drawn increasing attention for energy saving and latency reduction. However, due to massive connectivity, assigning different encoders, or pilots, is prohibitive for the tremendous number of users. To mitigate this issue, unsourced random access (URA), initially introduced by Polyanskiy in \cite{yury2017isit},  has emerged as a promising and pragmatic communication paradigm for handling massive uncoordinated users through a shared common codebook. In this way, the receiver is tasked to recover the set of messages up to permutations regardless of user identity.

\par Since the publication of \cite{yury2017isit}, numerous sophisticated techniques have been developed to approach the bound \cite{Vem2019TCOM,li2020gc,Amal2020TIT,Feng2021TIT,li2022JSAC,xie2022TCOM,Gkag2022SPAWC,gao2022TIT}. However, these approaches generally only consider the synchronous scenario. In reality, due to varying transmission distances and a lack of tight frequency synchronization, the received signals may be affected by  timing and frequency offsets (TO and FO, abbr. TFO), leading to phase rotations in frequency and time domains, respectively. To the best of our knowledge, most current research focuses on asynchronous scenarios with only TO \cite{yury2019ACSSC,Amal2019ICASSP,Decu2022GC}. Nevertheless, as an inevitable scenario, a TFO-coupled system puts higher requirements for the receiver design, rendering existing works unsuitable. Another line of work focuses on the GF-RA in the presence of both TO and FO \cite{sun2022TWC,yuan2021VTC}. Non-orthogonal pilots are utilized in \cite{sun2022TWC} in the MIMO-OFDM system, and a structured generalized approximate message passing (S-GAMP) is leveraged to estimate the TFO and channel jointly. While \cite{yuan2021VTC} unveils that the orthogonal pilots separate the multi-user interference, thus facilitating the TFO compensation.

\par Nonetheless, the pilot collision has been a bottleneck in URA due to the limited orthogonal space. To cope with this issue, most works resort to the non-orthogonal codebook \cite{Vem2019TCOM,li2020gc,Amal2020TIT,Feng2021TIT,li2022JSAC,xie2022TCOM,Gkag2022SPAWC}, which results in unmanageable dimensions, causing unsatisfactory complexity and residual user collisions. An energy detection-based scheme \cite{li2022JSAC} resolves the collision by data retransmission, compromising the efficiency due to interaction. \cite{Ahma2022isit} leverages multistage orthogonal pilots to reduce the collision density, while the channels are estimated individually at each stage, leading to suboptimality. In general, existing works primarily focus on the research of URA, either in synchronous scenarios or asynchronous with only TO and collision-free. In contrast, this paper aims to achieve reliable communication in both time and frequency asynchronous URA with user collisions. The main contributions are summarized as follows.
 
\par This paper is the first attempt of the study of collision resolution for MIMO massive URA in the asynchronous OFDM system with both TO and FO. Specifically, unlike existing codebook quantization-based approaches \cite{Amal2019ICASSP,sun2022TWC}, we directly conduct channel estimation (CE) by a minimum-mean squared error (MMSE) estimator to obtain the coarse channel coupled with the phase rotation caused by TFO. To reduce collision density, we utilize multistage orthogonal pilots and formulate an optimization problem to minimize the MSE of TFO-coupled channels across these stages. Motivated by the greedy algorithm and tree code proposed in \cite{Amal2020TIT}, we propose a novel graph-based scheme to iteratively reconstruct channels and resolve collisions jointly among multiple stages with manageable complexity, which is further generalized to the asynchronous case for TFO estimation and rotation compensation. Moreover, to reduce possible quantization errors, we establish a constellation-aided TFO correction algorithm by leveraging the geometric characteristics of the signal constellation. Exhaustive simulations demonstrate the remarkable performance advantages in CE and data decoding with substantial complexity reduction compared to the counterparts. 

\par \textit{Notation:} Throughout this paper, scalars, column vectors, and matrices are denoted by lowercase, bold lowercase, and bold uppercase, respectively. The transpose, conjugate, and conjugate transpose operations are signified by $\left( \cdot\right)^T, \left( \cdot\right)^*, \left( \cdot\right)^H$, respectively. $\left\| \mathbf{x} \right\|_p$ and $\left\| \mathbf{A} \right\|_F$ are the standard $l_p$ and Frobenius norms, respectively.  $\mathcal{R}(\cdot)$ denotes the real part of a complex value. $diag \left\lbrace \mathbf{d} \right\rbrace $ denotes the diagonal matrix with the vector $\mathbf{d}$ being the diagonal items. $\odot$ denotes the element-wise multiplication of two vectors or matrices.
%$\otimes$ represents the tensor product of matrices. 

\section{Problem Formulation} \label{sec-2}

% 前面两段有所重复，应当考虑整合

% \par Like many previous contributions, the user's message is portioned into two parts. Information in the first part is embedded in the codeword by a sparse regression code (SPARC) encoder, of which the interleaving pattern for the second part is also determined. While the message in the second part is coded by LDPC codes, then zero-padded, interleaved, and modulated in a sequential manner, which is so-called the interleave-division multiple access (IDMA). For the detailed encoding illustration, we refer the readers to \cite{li2022JSAC}. Additionally, in order to reduce the user collision density during the SPARC encoding, each user transmits multiple codewords picked from the codebook. Moreover, the tree coding scheme \cite{Amal2020TIT} is leveraged as an auxiliary method to recombine the messages embedded in multiple stages. 

\par Consider the uplink of a single-cell MIMO cellular network comprised of $K_{tot}$ single-antenna users, served by a base station (BS) equipped with $M$ antennas. Due to the sporadic traffic, a small set of $K_a$ users denoted by $\mathcal{K}_a$, with $K_a \ll K_{tot}$, are active to access the BS in a given transmission slot, each transmitting $B$ bits of information through $L$ time and frequency resources. We assume that $S$ out of total $N_c$, $S \ll N_c$, subcarriers are assigned for each user to transmit either pilot or data in each OFDM symbol. \footnote{In this case, a narrowband OFDM system is considered and the frequency-domain channel is modeled to be flat among subcarriers \cite{sun2022TWC}.} The overall transmission occupies $T$ OFDM symbols with $T_p$ and $T_d$ of them for pilots and data, respectively.
\begin{figure*}[t]
%	\vspace{-0.2cm}
	\centerline{\includegraphics[width=0.72\textwidth]{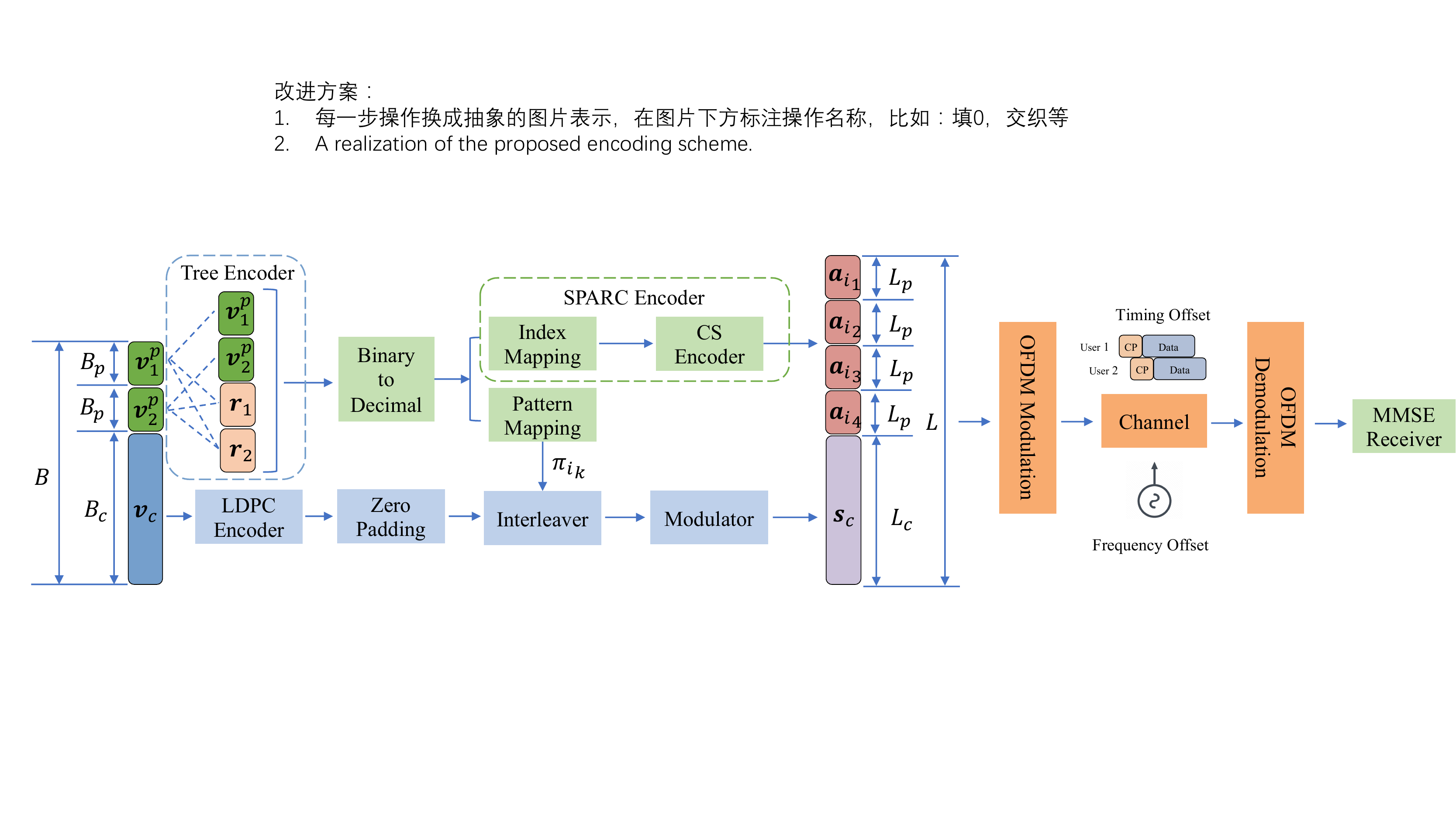}}
	\captionsetup{font=small}
	\caption{A realization of the overall encoding scheme and the proposed receiver design.}
	\label{pic-1}
	\vspace{-0.5cm}
\end{figure*}
\par A realization of the overall encoding scheme and receiver design is illustrated in Fig. \ref{pic-1}. Like many previous contributions, the user's information is first portioned into two parts, with $ \mathbf{v}_1^p, \mathbf{v}_2^p \in \left\lbrace 0,1 \right\rbrace^{B_p\times1} $ being the first part and $\mathbf{v}_c\in \left\lbrace 0,1 \right\rbrace^{B_c\times1}$ being the second part, which are referred to as the preamble and LDPC parts hereinafter, respectively. Specifically, multiple codewords are transmitted to reduce the user collision density in the preamble parts. To begin with, the tree encoder is leveraged as an auxiliary method to facilitate data splicing by generating two pieces of parity check data appended to the preamble, namely $\left\lbrace \mathbf{r}_1, \mathbf{r}_2 \right\rbrace$, where the detailed coding process can be found in \cite{Amal2020TIT}. Consequently, four pieces of data (each with length $B_p$) are then sent to the sparse regression code (SPARC) encoder to pick the corresponding codewords $\left\lbrace \mathbf{a}_{i_1} ,\cdots, \mathbf{a}_{i_4}\right\rbrace $ from the codebook $\mathbf{A} = \left[ \mathbf{a}_1, \mathbf{a}_2, \cdots, \mathbf{a}_{N} \right] \in \mathbb{C}^{L_p \times N}$ with $N=2^{B_p}$ and $L_p=S$. $\left\lbrace i_1, \cdots, i_4 \right\rbrace $ are not only the decimal representation (plus one) of the four pieces of data $\left\lbrace \mathbf{v}_1^p, \mathbf{v}_2^p, \mathbf{r}_1, \mathbf{r}_2  \right\rbrace$, respectively, and also determines the interleaving pattern of the data in LDPC phase. While in the LDPC part, the message is coded by LDPC codes, then zero-padded, interleaved, and modulated in a sequential manner, which is so-called the interleave-division multiple access (IDMA) \cite{ping2006twc}. For the detailed encoding illustration, we refer the readers to \cite{li2022JSAC}.

\par Due to the transmission delay and variation of oscillators, the received signal of user $k$ will suffer from the impact of TO and FO, denoted by $\tau_k$ and $\epsilon_k$, respectively. We assume that $\tau_k$ represents the residual TO after a closed-loop synchronization mechanism \cite{Decu2022GC} such that $\tau_k$ is no more than the cyclic prefix (CP) length $N_{CP}$. After removing the CP, the $t$-th OFDM symbol in the frequency domain can be expressed as
\begin{equation}
	\begin{aligned}
		\mathbf{Y}^t &= \sum\limits_{k=1}^{K_{a}}{ \mathbf{F_s} \mathbf{D}_{\epsilon_k}^t  \left(   \mathbf{I}_{N_c}\right) _{\tau_k}  \mathbf{F}_\mathbf{s}^H \mathbf{A}\mathbf{c}_k^t \mathbf{h}_k^T} + \mathbf{Z} \\
							   &= \sum\limits_{k=1}^{K_{a}}{ \mathbf{P}_{\epsilon_k}^t  \mathbf{P}_{\tau_k} \mathbf{A}\mathbf{c}_k^t \mathbf{h}_k^T} + \mathbf{Z}, t \in \left[ 1: T_p \right]
	\end{aligned} \label{equ-1}
\end{equation}
where $\mathbf{F_s} \in\mathbb{C}^{S\times N_c}$ is the partial matrix composed of $S$ row vectors extracted from the $N_c$-point discrete Fourier transform (DFT) matrix $\mathbf{F}$ indexed by the vectors $\mathbf{s}$, where $\mathbf{s} = \left[n_1,n_2, \cdots, n_S \right]^T\in\mathbb{Z}^{S\times1}$ is the vector of the subcarrier indices. $\mathbf{c}_k^t\in \left\lbrace 0,1 \right\rbrace^{N\times 1}$ is the binary selection vector of user $k$ at the $t$-th slot, which is all zero but a single one at $i_k$, the decimal representation (plus one) of the $B_p$-bit message produced by user $k$. In a narrowband system, the channel vector $\mathbf{h}_k  \in \mathbb{C}^{M\times1}$ is block fading and assumed to be independent and identically distributed (i.i.d.) with zero mean and variance $\sigma_h^2$. $\mathbf{Z}\in\mathbf{C}^{L_p\times M}$ is the addictive white Gaussian noise (AWGN) with each component distributed as $\mathcal{CN}\left(0, \sigma_n^2 \right)$.  $\mathbf{D}_{\epsilon_k}^t =\phi^t diag\left(1, \omega,\cdots,\omega^{N_c-1}  \right)\in\mathbb{C}^{N_c\times N_c}$ denotes the phase shift caused by $\epsilon_k$, with $\omega = e^{{j2\pi \epsilon_k}\slash{N_c}}$. While $\phi^t = \omega^{N_{cp}+(t-1)(N_{cp}+N_c)}$ represents the phase shift accumulated to the $t$-th symbol. $\left(\mathbf{I}_{N_c}\right) _{\tau_k}$ is a left-cyclically shifting matrix of $\mathbf{I}_{N_c}$ with $\tau_k$ units. The phase shift matrices $\mathbf{P}_{\epsilon_k}^t$ and $\mathbf{P}_{\tau_k}$ are respectively given by
\begin{align}
	\mathbf{P}_{\epsilon_k}^t &=   \mathbf{F}_{\mathbf{s}}  \mathbf{D}_{\epsilon_k}^t \mathbf{F}_{\mathbf{s}}^H = 	\phi^t \left[\mathbf{P} \right]_{\mathbf{s}\times\mathbf{s}} \\
	\mathbf{P}_{\tau_k} &  = \mathbf{F}_{\mathbf{s}}  \left(\mathbf{I}_{N_c}\right) _{\tau_k} \mathbf{F}_{\mathbf{s}}^H = diag \left\lbrace \bm{\psi} \right\rbrace 
\end{align}
where $\bm{\psi} = \left[ \psi^{1-n_1} \!, \!\cdots\!,\! \psi^{1-n_S} \right]^T \! \in \! \mathbb{C}^{S\times 1}, \psi = e^{j2\pi \tau_k \slash N_c}$ and 
\begin{equation}
	\mathbf{P} = \left[\!\!\!              %左括号
	\begin{array}{cccc}   %该矩阵一共3列，每一列都居中放置
		P(\epsilon_k) & P(1+\epsilon_k) & \cdots & P(N_c \!-\!1 \!+\!  \epsilon_k)\\  %第一行元素
		P(N_c \!-\!1 \!+\!  \epsilon_k) &  P(\epsilon_k) & \cdots & P(N_c \!-\! 2 \!+\! \epsilon_k)\\  %第二行元素
		\vdots &  \vdots &  \ddots &  \vdots \\
		P(1\!+\! \epsilon_k) & P(2\!+\!\epsilon_k)  & \cdots & P(\epsilon_k)
	\end{array} \!\!\!\right]. \notag
\end{equation}
$\left[\mathbf{P}\right] _{\mathbf{s}\times \mathbf{s}} \in\mathbb{C}^{S\times S}$ denotes the sub-matrix extracted from $\mathbf{P}\in \mathbb{C}^{N_c\times N_c}$ with the rows and columns indexed by the vector $\mathbf{s}$, and $P(\epsilon_k) = \frac{\sin \pi \epsilon_k}{N_c \sin \left( \pi \epsilon_k\slash N_c\right) } e^ {j  \pi \epsilon_k (N_c-1)\slash N_c }$. Note that $\epsilon_k$ can be controlled within a small range by detecting the downlink synchronization signals in practical communications, such as LTE or 5G NR \cite{sun2022TWC}. Consequently,  for the sake of model tractability and algorithm illustration, $\mathbf{P}^t_k \triangleq \mathbf{P}_{\epsilon_k}^t \mathbf{P}^t_{n_k} \in\mathbb{C}^{S\times S}$ can be simplified to a diagonal matrix with $\mathbf{p}_{k}^t  = \phi^t \omega^{(N_c-1)\slash 2} \left[\psi^{1-n_1}, \cdots, \psi^{1-n_S}  \right]^T \in\mathbb{C}^{S\times 1}$ denoting the diagonal elements. As such, Eq. \eqref{equ-1} can be rewritten as
\begin{equation}
	\mathbf{Y}^t  = \sum\limits_{k=1}^{K_{a}}{\left( \mathbf{A}\mathbf{c}_k^t\right) \odot\mathbf{p}_k^t\mathbf{h}_k^T +\mathbf{Z}  = \left(\mathbf{A}\mathbf{C}^t\right) \odot\mathbf{P}^t\mathbf{H}+\mathbf{Z}} \label{equ-4-add}
\end{equation}
where $\mathbf{C}^t = \left[\mathbf{c}_1^t,\cdots,\mathbf{c}_{K_a}^t\right]\in\left\lbrace 0,1\right\rbrace^{N\times K_a}$ denotes the binary selection matrix. $\mathbf{P}^t = \left[\mathbf{p}_1^t,\cdots,\mathbf{p}_{K_a}^t\right]\in\mathbb{C}^{S\times K_a}$ is the phase rotations. $\mathbf{H} = \left[\mathbf{h}_1,\cdots,\mathbf{h}_{K_a}\right]^T\in\mathbb{C}^{K_a\times M}$ is the channels. 

%\par We note that a comparable encoding scheme based on multistage pilots has been proposed in \cite{Ahma2022isit}. However, a fundamental difference is that in \cite{Ahma2022isit}, CE is conducted  independently at each stage without a collision resolution scheme design. In contrast, the proposed method involves jointly reconstructing the channel and resolving collisions across multiple stages, leading to improved estimation accuracy. Additionally, we generalize the proposed algorithm to the asynchronous scenario with both TO and FO existing. Further details of the proposed scheme are provided in \ref{Graph}. 

\section{Proposed Scheme}

%\par We proceed with the illustration of the overall receiver design. We start with introducing the CE method, then the graph-based scheme, and finally, the TFO correction algorithm.

\subsection{Activity Detection and CSI Acquisition}

\par To sufficiently reduce the multi-user interference and separate TFO phase rotations, we resort to the extremely sparse orthogonal pilot (ESOP), or unit matrix, as the codebook, i.e., $\mathbf{A} =\left[\mathbf{e}_1,\mathbf{e}_2,\cdots,\mathbf{e}_{N} \right]\in\left\lbrace0,1 \right\rbrace ^{L_p\times N}$ with $N=L_p$, where $\mathbf{e}_n\in\left\lbrace 0,1 \right\rbrace ^{L_p\times 1} $ is composed a single one in the $n$-th item and zeros elsewhere. As such, Eq. \eqref{equ-4-add} can  be simplified to 
\begin{equation}
	\mathbf{Y}^t  = \mathbf{C}^t \odot\mathbf{P}^t\mathbf{H}+\mathbf{Z}.
\end{equation}
Let $\mathbf{G}^t \triangleq  \mathbf{C}^t \odot\mathbf{P}^t\mathbf{H} = \left[\mathbf{g}^t_{1}, \cdots, \mathbf{g}^t_{N}\right]^T  \in \mathbb{C}^{N\times M}$, which is row-sparse and can be obtained by the MMSE estimator as
\begin{equation}
	\widehat{\mathbf{G}}^t = \mathbf{A}^H\left(\mathbf{A}\mathbf{A}^H + \sigma_n^2 \mathbf{I}_S \right)^{-1} \mathbf{Y}^t \triangleq {\mathbf{Y}^t}\slash{\sigma_n^2}.
\end{equation}
$\widehat{\mathbf{G}}^t  = \left[\hat{\mathbf{g}}_{1}^t,\cdots, \hat{\mathbf{g}}_{N}^t\right]^T \in \mathbf{C}^{N\times M}$ with $\hat{\mathbf{g}}_n^t$ given by
\begin{equation}
	\hat{\mathbf{g}}_{n}^t = \sum\nolimits_{k\in\mathcal{K}_n}{\mathbf{p}^t_{k}(n)\mathbf{h}_k} + \mathbf{z}
\end{equation}
where $\mathcal{K}_n$ is the set of users choosing the codeword $\mathbf{e}_n$, which may be a \textit{zeroton,} \textit{singleton}, or \textit{multiton,} referring to as no user, single user, or multiple users (collision case) choosing $\mathbf{e}_n$. Intuitively, $\mathbf{e}_n$ and $\mathbf{g}^t_{n}$ can be obtained by the hard decision on the row energy of $\widehat{\mathbf{G}}_t$. However, it is non-applicable to the collision case with the superimposed of channels,  which can be well addressed according to the following algorithm.

\subsection{Graph-Based Channel Reconstruction and Collision Resolution} \label{Graph}

%Note that, due to the limited tree coding constraints, the validity of the edges cannot be guaranteed, i.e., both valid and invalid edges exist. Correspondingly, tree coding is employed as an auxiliary method for the trade-off of decoding complexity and unguaranteed accuracy. Furthermore, the proposed GB-CR$^2$ algorithm assigns weights to all edges based on the channel MSE across multi-segments and aims to minimize the overall MSE of the graph in order to verify the validity of edges and reconstruct users' channels. We provide an example of a graph in Fig. \ref{pic-2}.

%Note that the proposed graph is similar to a bipartite graph, but with a key difference: due to the presence of collisions, nodes in the graph are not necessarily connected to only one node. Therefore, it is not feasible to simply separate the nodes into one-to-one connections. 

\par Firstly, we review the two key differences in the proposed scheme distinguished from \cite{Ahma2022isit,yuan2021VTC} as follows:
\begin{itemize}
	\item Based on the CE results obtained from multistage pilots, the collided channels are required to be separated and reconstructed. 
	\item Due to the IDMA framework in the LDPC part, the data embedded in multiple pilot segments must be spliced together to recover the interleaving pattern. 
\end{itemize}
\par Based on the above two characteristics, we summarize the algorithm design into two specific tasks: 1) Cross-segment splicing of user data and 2) channel reconstruction associated with collision resolution and TFO compensation. We propose the\textit{ Graph-Based Channel Reconstruction and Collision Resolution (GB-CR$^2$)} algorithm for these two missions. We begin the illustration with the definition of associated items. 
 \begin{itemize}
 	\item \textit{Node:}  The user's data at each stage. Nodes are isolated within stages but connected across stages through edges.
 	\item \textit{Edge:}  An edge represents a \textit{possible} connection between nodes, which are initialized in the tree decoding process. 
 	\item \textit{Weight:} An edge's weight reflects its credibility, assigned by the GB-CR$^2$ algorithm according to the channel MSE.
 \end{itemize}
\par
\begin{figure}[htbp]
 	\vspace{-0.2 cm}
 	\centerline{\includegraphics[width=0.38\textwidth]{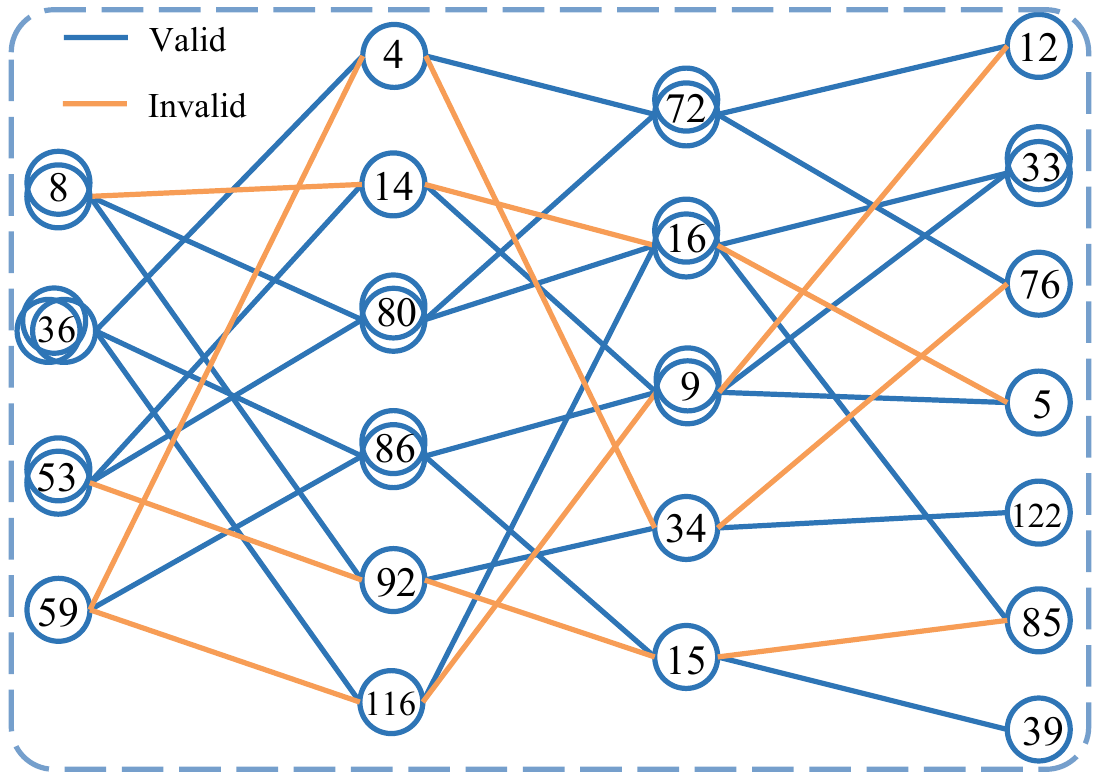}}
 	\captionsetup{font=small}
 	\caption{An example of the proposed graph with $K_a=8$ and four stages, with both valid (colored blue) and invalid (colored orange) edges. Nodes with multiple circles correspond to collision cases.}
 	\vspace{-0.2 cm}
 	\label{pic-2}
 \end{figure}

 We further introduce variables $\left\lbrace N_i, \mathbf{v}_i, \mathbf{D}^i \right\rbrace_{i=1}^{T_p} $ and the set $\mathcal{P}$ to characterize the graph. $N_i $ denotes the number of nodes in the $i$-th stage, and $\mathbf{v}_i\in \mathbb{Z}^{N_i\times1}$ is the vector of users' messages (nodes) sorted in an ascending order. $\mathbf{D}^i\in \left\lbrace0,1 \right\rbrace ^{K_a \times N_i}$ is a binary \textit{selection matrix} with its element $\mathbf{D}^i(m,n) = 1$ if user $m$ selects the $n$-th message or zero otherwise, and we have 
 \begin{align}
	&\sum \nolimits_{n=1}^{N_i}{\mathbf{D}^i\left(:,n\right) } = \mathbf{1}^{K_a\times1}\\
	& \sum \nolimits_{k=1}^{K_a}{\mathbf{D}^i\left(k,:\right) } \geq \mathbf{1}^{1\times N_i}.
\end{align}
 $\mathcal{P}$ is the set of all existing paths on the graph generated by the tree decoder, i.e., a path $\left\lbrace \mathbf{v}_1(n_1), \mathbf{v}_2(n_2), \mathbf{v}_3(n_3), \mathbf{v}_4(n_4)\right\rbrace  \in \mathcal{P}$ if these nodes are connected together. Specifically, we provide an example of a graph in Fig. \ref{pic-2}, where $\mathbf{v}_1=\left[ 8,36,53,59\right]$ and $\mathbf{D}^1$ is given by
 \begin{equation}
 	\mathbf{D}^1 = \left[                  %左括号
 	\begin{array}{cccccccc}   %该矩阵一共3列，每一列都居中放置
 		1 & 1 & 0 & 0& 0 & 0& 0 & 0\\  %第一行元素
 		0 & 0 & 1 & 1& 1 & 0& 0 & 0\\  %第二行元素
 		0 & 0 & 0 & 0& 0 & 1& 1 & 0\\
 		0 & 0 & 0 & 0& 0 & 0& 0 & 1
 	\end{array}
 	\right]^T \! \! \! \! \in \left\lbrace0,1 \right\rbrace ^{K_a\times N_1}.
 \end{equation}
In Fig. \ref{pic-2}, $\left\lbrace 8, 80, 16, 33\right\rbrace \in \mathcal{P}$ since these nodes are connected and there is a path among them, and $\left\lbrace 8, 80, 16, 12\right\rbrace \notin \mathcal{P}$ since no path goes through the corresponding nodes.  

 \par  Note that both valid and invalid edges exist due to the limited tree coding constraints. Therefore, tree coding can only be an auxiliary method for the trade-off of decoding complexity and unguaranteed accuracy. While the validity of edges is further verified by the GB-CR$^2$ algorithm based on the MSE of channels across multi-segments.  Moreover,  the phase rotation coupled with $\mathbf{g}_{n}^t$ needs to be eliminated when calculating the path weight, since it varies on different subcarriers and symbols. 
 For tractability, we assume that the TO is integer-sampled within $\mathbf{d}=\left[1,2,\cdots,D \right]^T$ and FO is within $\mathbf{q}=\left[\epsilon^{(1)},\epsilon^{(2)},\cdots,\epsilon^{(Q)} \right]^T$, which is uniformly sampled from the range $\left[-\epsilon_{max}, \epsilon_{max}\right]$. Consequently, we are ready to introduce the overall optimization object as follows:
\begin{align}
	\widehat{\mathbf{\Theta}} = \arg \min _{\mathbf{\Theta}} &{\!\!\!\!\!\!\!\!\!\sum\limits_{1\leq i \leq j \leq T_p \atop \left\lbrace \mathbf{v}_1(n_1), \cdots, \mathbf{v}_{T_p}(n_{T_p})\right\rbrace \in \mathcal{P}} {\!\!\!\!\!\!\!\!\!\!\!\!\!\!\!\!\!\!\left\| \left( \widehat{\mathbf{G}}^i \left(\mathbf{v}_i(n_i), :\right) \! -\! \mathbf{D}^i\left(:,n_i\right)^T\mathbf{G}^i \right) \right. }} \notag  \\
	& {{\left. - \left( \widehat{\mathbf{G}}^j \left(\mathbf{v}_j(n_j), :\right) - \mathbf{D}^j\left(:,n_j\right)^T{\mathbf{G}}^j\right) \right\| _2^2}} \label{equ-10}
\end{align}
where $\mathbf{\Theta} = \left\lbrace K_a,\mathbf{H}, \mathbf{D}^i, \tau_k, \epsilon_k\right\rbrace$, $\mathbf{H} = \left[\mathbf{h}_1, \cdots, \mathbf{h}_{K_a} \right]^T, {\mathbf{G}}^i = \left[\mathbf{g}_{1}^i, \cdots, \mathbf{g}_{K_a}^i \right]^T \in\mathbb{C}^{K_a\times M}$ with $\mathbf{g}_{k}^i =  \mathbf{p}_{k}^i({\mathbf{v}_i}(n_i))\mathbf{h}_k$. Note that the optimization of $\mathbf{\Theta}$ in Eq. \eqref{equ-10} encompass a successive interference cancellation (SIC) process with the introduction of $\mathbf{D}^i$, i.e., the channels of collided nodes have been separated or eliminated to the greatest extent to obtain the smallest MSE on each edge. However, Eq. \eqref{equ-10} is a mixed integer nonlinear programming (MINLP) problem, which is NP-Hard. To solve this problem effectively, the proposed GB-CR$^2$ algorithm iteratively calculates the MSE of paths in the current graph. The MSE the path $p \in \mathcal{P}$ is
\begin{align}
	 MSE\left(p, \left\lbrace \widehat{\mathbf{G}}^i, \mathbf{v}_i\right\rbrace _{i=1}^{T_p}, {\mathbf{d}}(m), {\mathbf{q}}(n)\right) = \;\;\;\;\;\;\;\;\;\;\;\;\;\;\;\;\;\;\; \notag  \\
	 \sum\limits_{1\leq i \leq j \leq T_p}{ \left\| \frac{ \widehat{\mathbf{G}}^i \left(\mathbf{v}_i(n_i), :\right) }{\mathbf{p}_{m,n}^i(\mathbf{v}_i(n_i))} - \frac{ \widehat{\mathbf{G}}^j \left({\mathbf{v}}_j(n_j), :\right) }{\mathbf{p}_{m,n}^j(\mathbf{v}_j(n_j))} \right\| _2^2 } \label{equ-11}
\end{align}
with $ \left\lbrace \mathbf{v}_1(n_1), \cdots, \mathbf{v}_{T_p}(n_{T_p})\right\rbrace = p$, and  
\begin{equation}
	\mathbf{p}_{m,n}^i(k) = \psi^{1-\mathbf{s}(k)}  \omega^{(N_{cp}+N_c)i-\frac{N_c+1}{2}} \label{equ-12}
\end{equation}
with $\psi = e^{i2\pi \mathbf{d}(m)\slash N_c}$ and $\omega = e^{j2\pi \mathbf{q}(n)\slash N_c}$. Generally, we can restore a path by the minimum weight path searching, i.e.,  
\begin{equation}
	\left[ \hat{p}, \hat{\tau}_k, \hat{\epsilon}_k \right] = \arg \min_{p\in \mathcal{P}} {MSE\left(p, \left\lbrace \widehat{\mathbf{G}}^i, \mathbf{v}_i\right\rbrace _{i=1}^{T_p}, {\mathbf{d}}, {\mathbf{q}}\right)} \label{equ-13}
\end{equation}
where the TFO estimation can be simultaneously obtained by searching in $\mathbf{d}$ and $\mathbf{q}$ to minimize the path weight. The path $\hat{p}$ with the smallest MSE is regarded as the highest reliability and we further define that $\hat{p}$ is valid if the arbitrary nodes $\mathbf{v}_i(n_i), \mathbf{v}_j(n_j)$ of $\hat{p}$ satisfies that  
\begin{equation}
	\left\| \widehat{\mathbf{H}}^i\left( \mathbf{v}_i(n_i), :\right) - \widehat{\mathbf{H}}^j\left( \mathbf{v}_j(n_j), :\right) \right\|_2^2 < max \left\lbrace a,b\right\rbrace \label{equ-14}
\end{equation}
where $a=\left\|\widehat{\mathbf{H}}^i\left( \mathbf{v}_i(n_i), :\right)  \right\|_2^2, b=\left\| \widehat{\mathbf{H}}^j\left( \mathbf{v}_j(n_j), :\right) \right\|_2^2$, and $\widehat{\mathbf{H}}^i\left( \mathbf{v}_i(n_i), :\right) = { \widehat{\mathbf{G}}^i \left(\mathbf{v}_i(n_i), :\right) }\slash {\mathbf{p}_{m,n}^i(\mathbf{v}_i(n_i))}$. And we say the node $\mathbf{v}_i(n_i)$ of $\hat{p}$ is non-collided if
\begin{align}
	\left\| \widehat{\mathbf{H}}^i\left( \mathbf{v}_i(n_i), :\right) - \widehat{\mathbf{H}}^j\left( \mathbf{v}_j(n_j), :\right) \right\|_2^2 & \leq  \gamma \label{equ-15}
\end{align}
where $\mathbf{v}_j(n_j)$ is the node of $\hat{p}$ with the lowest channel energy and $\gamma$ is a predefined threshold. Let $\mathcal{V}_p$ denote the set of non-collided nodes of path $p$ (satisfying Eq. \eqref{equ-15}), then $\widehat{\mathbf{h}}_k$ is
\begin{equation}
	\widehat{\mathbf{h}}_k = 1\slash \left|\mathcal{V}_p \right|  \sum\nolimits_{\mathbf{v}_i(n_i) \in \mathcal{V}_p}{\widehat{\mathbf{H}}^i\left( \mathbf{v}_i(n_i), :\right)^T}. \label{equ-16}
\end{equation}
 which is then eliminated in the collided nodes. 
 \begin{figure}[htbp]
 	\vspace{-0.2cm}
 	\centerline{\includegraphics[width=0.45\textwidth]{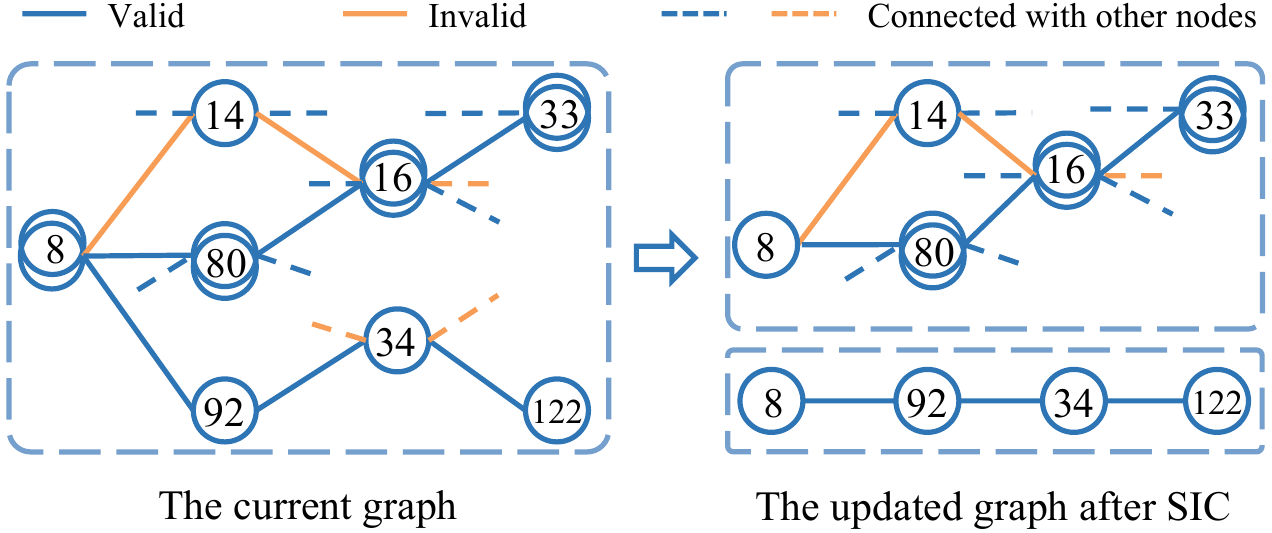}}
 	\captionsetup{font=small}
 	\caption{The SIC and pruning process on a sub-graph. The path $\left\lbrace 8,92,34,122\right\rbrace$ is extracted and then eliminated from the graph.}
 	\label{pic-3}
 	\vspace{-0.2cm}
 \end{figure}
 \par Once a path is picked up, the corresponding edges will be deleted in the graph, which goes smaller until there is no node or edge. We give a sub-graph in Fig. \ref{pic-3} to demonstrate this process, where the path $\left\lbrace 8,92,34,122\right\rbrace$ currently has the smallest MSE, thus being extracted and eliminated from the graph. The overall algorithm is summarized in Alg. \ref{alg-1}.  
 \begin{algorithm} [htpb]
 	\setstretch{1.15}
 	\caption{The Proposed GB-CR$^2$ Algorithm}
 	\label{alg-1}  
 	\begin{algorithmic}[1]	
 		\STATE {{\bf Input}: $\mathcal{P}$, $\gamma$, $\mathbf{d}$, $\mathbf{q}$, $\mathbf{v}_i, \widehat{\mathbf{G}}^i, \forall i\in [1:T_p]$}\\
 		\STATE{{\bf Initial}: $ \widehat{\mathcal{P}}= \emptyset$, $\widehat{K}_a=0$, $\widehat{\mathbf{H}}= [\;]$}
 		\STATE {{\bf Output}: {$\hat{\tau}_k$, $\hat{\epsilon}_k$, $ \widehat{\mathcal{P}}$, $\widehat{\mathbf{H}}, \forall k\in [1:\widehat{K}_a]$}}
 		\REPEAT
 		\STATE {{ \% Minimum Weight Path Searching}}\\
 		\STATE {Update: $\left(\hat{p},\hat{\tau}_k, \hat{\epsilon}_k\right)$ via \eqref{equ-13}, $\mathcal{P} \leftarrow \mathcal{P} - \hat{p}$}\\
 		\STATE {{\bf if} $\hat{p}$ satisfying \eqref{equ-14}}\\
 		\STATE {\quad Update: $\widehat{\mathbf{h}}_k$ via \eqref{equ-16}, $\widehat{\mathbf{H}}(k,:) \leftarrow \widehat{\mathbf{h}}_k^T$}\\
 		\STATE {\quad Update: $\widehat{K}_a \leftarrow \widehat{K}_a+1, \widehat{\mathcal{P}} \leftarrow \widehat{\mathcal{P}} \cup \hat{p} $}\\
 		\STATE {\quad \% Successive Interference Cancellation}\\
 		\STATE {\quad {\bf if} $\mathbf{v}_i(n_i)$ NOT satisfying \eqref{equ-15}}\\
 		\STATE{\quad \quad  $\widehat{\mathbf{G}}^i \left(\mathbf{v}_i(n_i), :\right) \!\leftarrow\! \widehat{\mathbf{G}}^i \left(\mathbf{v}_i(n_i), :\right) \!-\! \mathbf{p}_{m,n}^i(\mathbf{v}_i(n_i))\widehat{\mathbf{h}}_k^T$}\\
 		\STATE {\quad {\bf end}}\\
 		\STATE {{\bf end}}\\
 		\UNTIL {$\mathcal{P}=\emptyset$}
 	\end{algorithmic}  
 \end{algorithm}
 
\subsection{Constellation-Aided TFO Correction}

%\par The TFO value obtained in Alg. \ref{alg-1} is estimated based on the minimum MSE of multi-segment channels within the range of quantization values. However, due to the approximation error in Eq. \eqref{equ-12} and the limited observations of ESOP, the TFO estimation in Alg. \ref{alg-1} is not always optimal. In fact, the phase rotation caused by TFO also occurs in the data segment, which occupies more OFDM symbols than the pilots, and thus has more observations. Therefore, the data observations can be utilized to verify and correct the TFO estimation obtained in Alg. \ref{alg-1} to obtain a better estimation of TFO. The received $t$-th symbol, $t\in[T_p+1: T_p+T_d]$, in the LDPC part is

\par The TFO value obtained above is based on the minimum MSE of multistage channels within the range of quantization values. However, due to the approximation error in Eq. \eqref{equ-12} and the limited observations of ESOP, the TFO estimation in Alg. \ref{alg-1} is not always optimal. In fact, the phase rotation caused by TFO also occurs in the LDPC part, which occupies more OFDM symbols than the pilots, and thus has more observations. Therefore, the data can be utilized to correct the TFO estimation obtained in Alg. \ref{alg-1} to obtain a better one. The received $t$-th symbol, $t\in[T_p+1: T_p+T_d]$, is given by
\begin{equation}
	\mathbf{Y}^t = \sum\limits_{k=1}^{K_{a}}{\mathbf{P}_{k}^t  f\left(\mathbf{s}_k^c \right)  \mathbf{h}_k^T} + \mathbf{Z}
\end{equation}
where $f(\mathbf{s}_k^c)$ refers to the encoding process described in Section \ref{sec-2} on $\mathbf{s}_k^c$, the $k$-th user's modulated symbol. After the MMSE estimation, the received signal can be expressed as
\begin{equation}
	\widehat{\mathbf{X}} = \widehat{\mathbf{H}}^*\left( \widehat{\mathbf{H}}^T \widehat{\mathbf{H}}^* + \sigma_n^2 \mathbf{I}_M \right)^{-1} \Big( \mathbf{Y}^t\Big)^T.
\end{equation}
Consequently, the data after TFO compensation is $\widehat{\mathbf{s}}_k^c(s) = f^{-1}\left(\widehat{\mathbf{X}}(k,s) \right)\slash \mathbf{p}_{m,n}^t(s)$, of which the constellation should be similar to $\mathbf{s}_k^c$. However, due to the sub-optimal estimation of TFO in Alg. \ref{alg-1}, the residual TFO estimation error will cause a slight phase rotation on $\widehat{\mathbf{s}}_k^c$. Therefore, we can generate a list of $N$ TFO samples as coarse estimations by the GB-CR$^2$ algorithm, and the optimal TFO estimation can be obtained by finding the one with the minimal phase rotation on the data. Here, we introduce variable $\rho$ to describe the \textit{compensation degree}. For BPSK modulation \footnote{ We note that relatively low-order modulations are preferred to exploit the geometric characteristics of the constellation, and the algorithms can be easily extended to QPSK case.}, $\rho$ is given by
\begin{equation}
	\rho_{k,n} = \sum\nolimits_{\mathcal{R}\left\lbrace \widehat{\mathbf{s}}_{k,n}^c \right\rbrace>0}{\widehat{\mathbf{s}}_{k,n}^c}- \sum\nolimits_{\mathcal{R}\left\lbrace \widehat{\mathbf{s}}_{k,n}^c \right\rbrace<0}{\widehat{\mathbf{s}}_{k,n}^c}. 
\end{equation}
\par
\begin{figure}[htpb]
	\vspace{-0.2cm}
	\centerline{\includegraphics[width=0.48\textwidth]{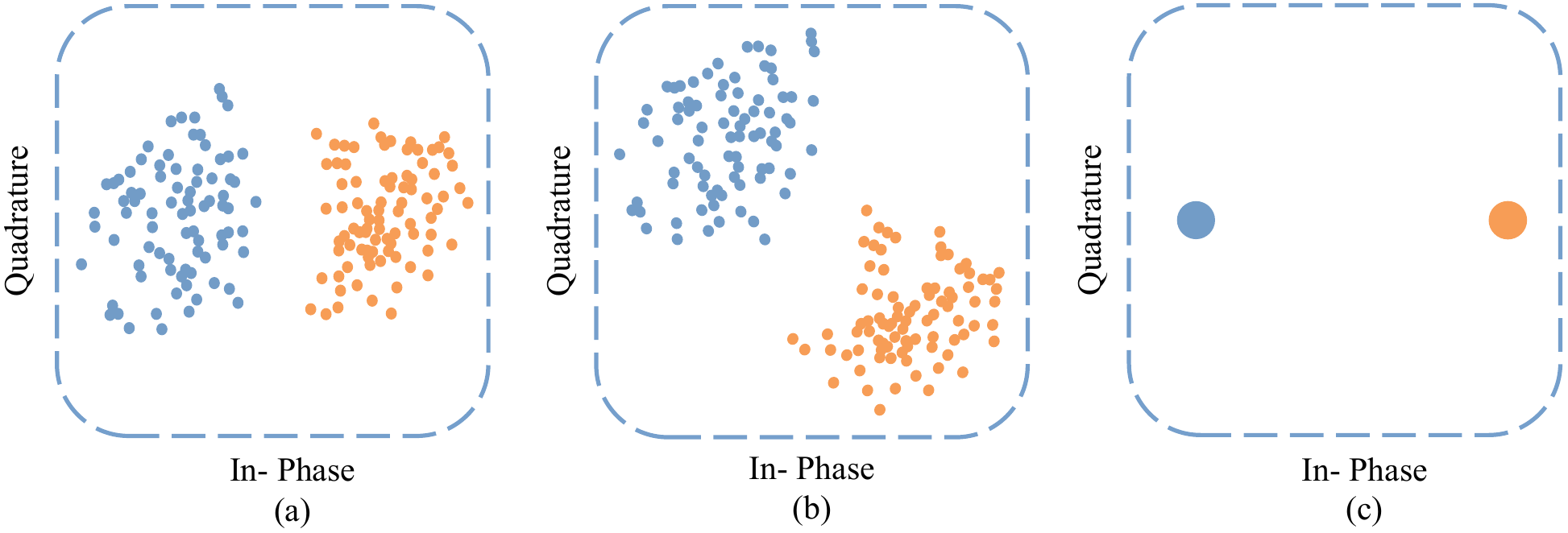}}
	\captionsetup{font=small}
	\caption{The phase rotation on the constellation of BPSK modulation. (a) $\widehat{\mathbf{s}}_{k}^c$ with a perfect rotation compensation. (b) Imperfect compensation on $\widehat{\mathbf{s}}_{k}^c$ with residual phase rotation and $\left\| \rho_a \right\|> \left\| \rho_b\right\|$. (c) The constellation of transmitted signal $\mathbf{s}_{k}^c$.}
	\label{pic-4}
	\vspace{-0.2cm}
\end{figure}

 Fig. \ref{pic-4} manifests the constellation of different phase compensation cases. Apparently, a more accurate TFO estimation leads to a more sufficient compensation of the phase ration, and thus the larger $\left\| \rho\right\|$ should be. As such, the optimal TFO estimation can be obtained by
\begin{equation}
	\left[\hat{\tau}_k, \hat{\epsilon}_k \right] = \arg \max_{\hat{\tau}_{k,n}, \hat{\epsilon}_{k,n}} {\left\| \rho_{k,n}\right\|}, \forall n \in [1:N]. 
	\vspace{-0.3cm}
\end{equation}
\par Moreover, as we will see shortly in Section \ref{sec-4}, an improved CE results can be obtained by plugging the estimated TFO results to the GB-CR$^2$ algorithm and re-estimating the channel with known TFO values. The subsequent LDPC decoding is by the belief propagation (BP) iterative structure. The complexity of the proposed algorithm in the preamble part is $\mathcal{O}(NL_p^2+MDQ)$, while that of the S-GAMP algorithm proposed in \cite{sun2022TWC} is $\mathcal{O}(M^2ND^2Q^2)$. Moreover, the quantization space of GB-CR$^2$ is $DQ$, while that of S-GAMP is $NDQ$. Altogether, compared to the counterpart, the proposed algorithm exhibits substantial computational complexity reduction. 

\section{Numerical Results}\label{sec-4}

\par We conduct numerical experiments to evaluate the performance of the proposed scheme compared to state-of-the-art schemes. The S-GAMP algorithm \cite{sun2022TWC} is utilized as the benchmark for the CE performance, evaluated by the normalized MSE (NMSE), i.e., NMSE $= \left\| \mathbf{H} - \underline{\mathbf{H}} \right\|_F^2 \slash \left\|\mathbf{H} \right\|_F^2$, where $\underline{\mathbf{H}}$ denotes the matrix $\widehat{\mathbf{H}}$ arranged by the indexes aligned with $\mathbf{H}$. While the FASURA scheme \cite{Gkag2022SPAWC} is employed as the baseline for the block error rate (BLER) performance, evaluated by the probability of misdetection $P_{md}$ and false alarm $P_{fa}$ given by
\begin{align}  
	 P_{md} &= \frac{1}{{{K_a}}}\sum\nolimits_{k \in {\mathcal{K}_a}} {P\left( {{\bm{v}_{{k}}} \notin \mathcal{L}} \right)}  \\
	 P_{fa} &= \frac{{\left|  {\mathcal{L}\backslash \left\{ {{\bm{v}_{{k}}}:k \in {\mathcal{K}_a}} \right\}} \right|}}{{\left| \mathcal{L}\right|}}
\end{align} 
where $\mathcal{L}$ is the recovered message list. The system's energy per bit is defined as ${E_b\slash N_0} = LP\slash BN_0$, where $L$ is the total channel uses, and $P$ is the symbol power. Specifically, we choose $B=100$ message bits and $L=3200$ channel uses. Two pieces of messages with a length $B_p = 7$ in the preamble part are encoded into four codewords, with each the length $L_p = 128$. While the rest $B_c=86$ message bits are encoded by an $\left(\underline{L}_c, B_c\right)$ LDPC code with $\underline{L}_c=200$, and then zero-padded, interleaved, and BPSK-modulated sequentially, resulting in total $L_c=2688$ channel uses. For the OFDM settings, $N_c=1024$, $N_{cp}=72$, $S=128$, $T_p=4$, and $T_d=21$. The subcarrier indices $\mathbf{s} = \left[ 1, 3, \cdots, 255\right]$ and are set to be identical to all users.  For the TFO settings, aligned with \cite{sun2022TWC}, the maximum TO and FO are $D=9$ and $\epsilon_{max}=0.0133$, respectively, with both the quantization number $D=Q=9$. 
\begin{figure}[htbp]
	\centerline{\includegraphics[width=0.38\textwidth]{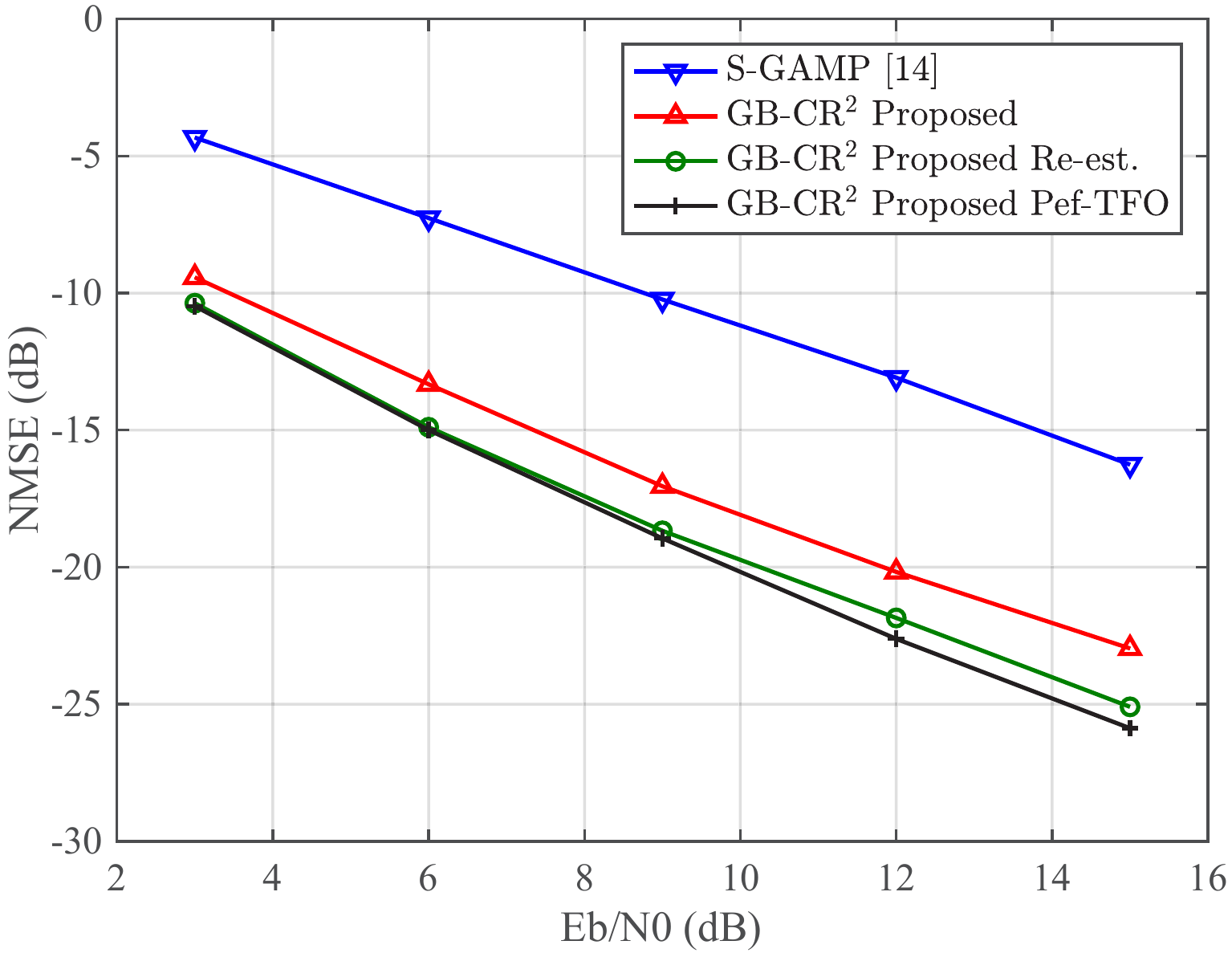}}
	\captionsetup{font=small}
	\caption{The CE performance comparison in various $E_b\slash N_0$. Parameter settings:  $M=16$, $K_a=60$, $L_p=N=512$ for S-GAMP.}
	\label{pic-5}
%	\vspace{-0.3cm}
\end{figure}
\par We depict the CE performance comparison versus the SNR in Fig. \ref{pic-5}, where the user collision is not considered for fair comparison, due to the non-applicability of S-GAMP in such case, although our scheme can handle it. In Fig. \ref{pic-5}, the tag `Re-est.' denotes the channel re-estimation with estimated TFO, while `Pef -TFO' refers to the CE under perfect TFO, a lower bound for evaluating the performance. As depicted in Fig. \ref{pic-5}, the proposed GB-CR$^2$ algorithm significantly exhibits an overall $8.4$ dB enhancement in the CE performance compared with  S-GAMP, with only a $0.3$ dB loss to the lower bound with known TFO. The substantial gain may result from the more accurate TFO estimation in the proposed scheme by jointly utilizing multi-segment channel observations. 
\begin{figure}[htbp]
	\centerline{\includegraphics[width=0.39\textwidth]{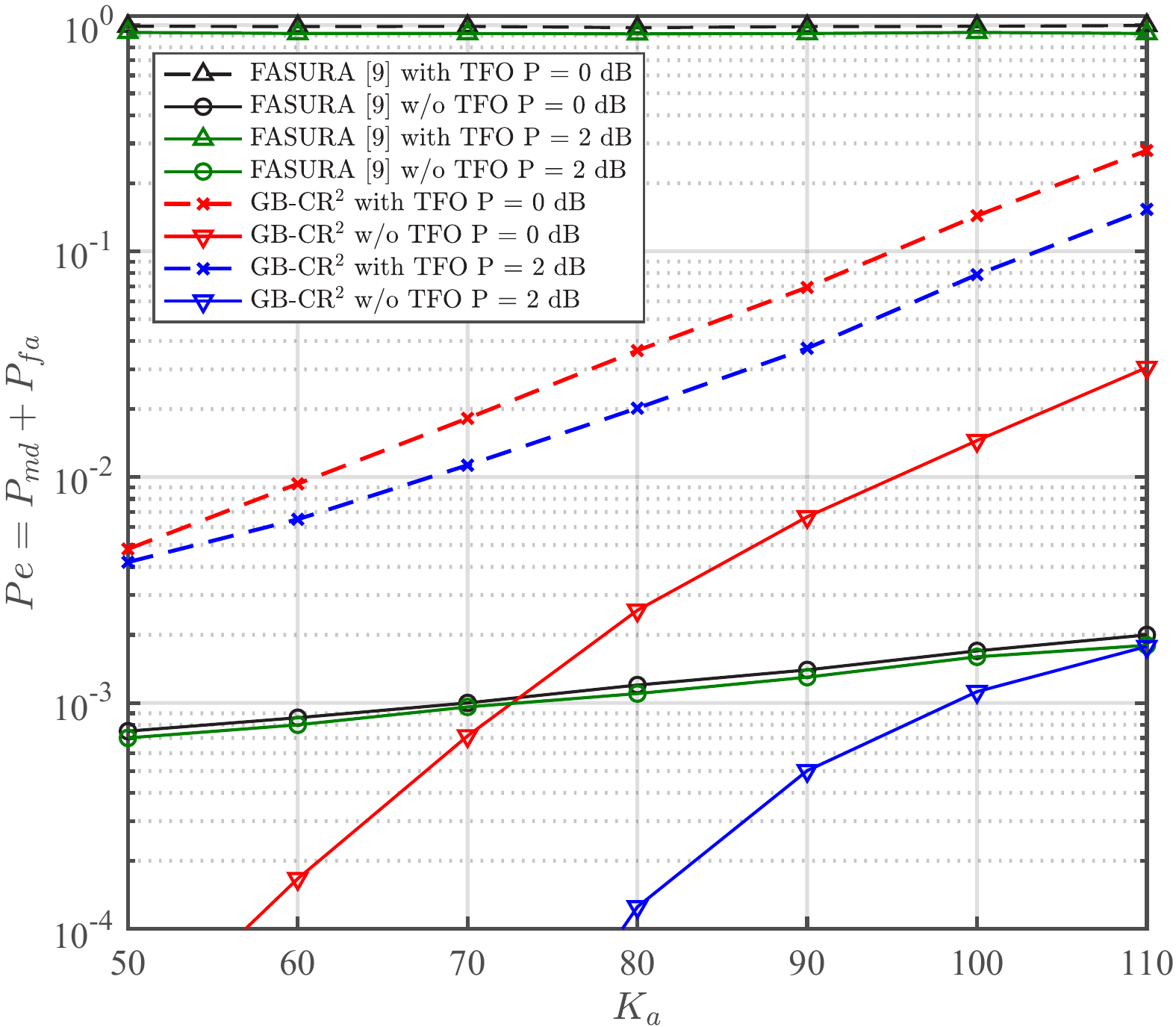}}
	\captionsetup{font=small}
	\caption{The BLER performance comparison of different schemes in $M=16$ and various $K_a$ and $P$.}
	\label{pic-6}
	\vspace{-0.3cm}
\end{figure}
\par Regarding that currently there is little work considering TFO in URA, both TFO-existing and TFO-free scenarios are considered in the simulation. We depict the curve of $P_e$ versus $K_a$ and $P$ in Fig. \ref{pic-6} to evaluate the BLER performance of the proposed scheme compared to FASURA with the parameters aligned with \cite{Gkag2022SPAWC}. Specifically, we utilize the non-orthogonal Gaussian codebook in the TFO-free scenario, where $B_p = 12$ and $B_c = 76$ with the other settings unchanged. Evidently, FASURA exhibits poor performance in the presence of TFO, while the proposed GB-CR$^2$ algorithm can handle it with an acceptable BLER, of which the performance is further improved by $2.5$ dB with the transmitting power from $0$ to $2$ dB. Convincingly, the comparison under the TFO-free scenario is also given in Fig. \ref{pic-6} for completeness. It is shown that GB-CR$^2$ outperforms FASURA in the regime of $K_a<70$ with $P=0$ dB, and this superiority is further generalized to $K_a<110$ in $P=2$ dB. On the contrary, the user collisions have become the bottleneck for FASURA, since there is no consideration for the collision resolution in this scheme. Altogether, the proposed algorithm outperforms state-of-the-art scheme with the superiority of addressing the TFO disturbance and accommodating more users with the rising transmitting power.

\section{Conclusion}
\par This paper considers the asynchronous MIMO URA system with the existence of both TFO and non-negligible user collisions. We propose a novel algorithmic solution called GB-CR$^2$ to reconstruct channels, compensate both TO and FO, and resolve collisions. We further improve the performance by leveraging the geometric characteristics of signal constellations. The pivotal of GB-CR$^2$ is to formulate the multi-stage transmission to a bipartite graph and iteratively recover the key parameters without requiring the quantization codebook with a large size. Our numerical results manifest the superiority of the proposed algorithm compared to the counterparts in terms both of CE and data recovery with substantial complexity reduction. 

% 未来方向可以是频选信道


\begin{thebibliography}{00}
	
	 \bibitem{wu2020wcm}
	Y. Wu, X. Gao, S. Zhou, W. Yang, Y. Polyanskiy, and G. Caire, ``Massive access for future wireless communication systems," \emph{IEEE Wireless Commun.}, vol. 27, no. 4, pp. 148-156, Aug. 2020.	
	
	 \bibitem{yury2017isit}
	Y.~{Polyanskiy}, ``A perspective on massive random-access," in \emph{Proc. IEEE Int. Symp. Inform. Theory (ISIT)}, Aachen, Germany, June 2017, pp. 2523--2527.
	
	  \bibitem{Vem2019TCOM}
	A. Vem, K. R. Narayanan, J. Chamberland, and J. Cheng, ``A user-independent successive interference cancellation based coding scheme for the unsourced random access Gaussian channel," \emph{IEEE Trans. Commun.}, vol. 67, no. 12, pp. 8258-8272, Dec. 2019.
	
	 \bibitem{li2020gc}
	T. Li, Y. Wu, M. Zheng, D. Wang, and W. Zhang, ``SPARC-LDPC coding for MIMO massive unsourced random access," in \emph{Proc. IEEE Globecom Workshops (GC Wkshps)}, Taipei, Taiwan, Dec. 2020, pp. 1-6.
	
	\bibitem{Amal2020TIT}
	V. K. Amalladinne, J. -F. Chamberland, and K. R. Narayanan, ``A coded compressed sensing scheme for unsourced multiple access," \emph{IEEE Trans. Inf. Theory}, vol. 66, no. 10, pp. 6509-6533, Oct. 2020.
	
	 \bibitem{Feng2021TIT}
	A. Fengler, S. Haghighatshoar, P. Jung, and G. Caire, ``Non-Bayesian activity detection, large-scale fading coefficient estimation, and unsourced random access with a massive MIMO receiver," \emph{IEEE Trans. Inf. Theory}, vol. 67, no. 5, pp. 2925-2951, May 2021.
			 
	\bibitem{li2022JSAC}
	T. Li et al., ``Joint device detection, channel estimation, and data decoding with collision resolution for MIMO massive unsourced random access,"  \emph{IEEE J. Sel. Areas Commun.}, vol. 40, no. 5, pp. 1535-1555, May 2022.
	
	\bibitem{xie2022TCOM}
	X. Xie et al., ``Massive unsourced random access: exploiting angular domain sparsity," \emph{IEEE Trans. Commun.} , vol. 70, no. 4, pp. 2480-2498, April 2022.	
		
	\bibitem{Gkag2022SPAWC}
	M. Gkagkos, K. R. Narayanan, J. -F. Chamberland, and C. N. Georghiades, ``FASURA: a scheme for quasi-static massive MIMO unsourced random access channels," in \emph{Proc. IEEE Workshop Signal Process. Adv. Wireless Commun. (SPAWC)}, Oulu, Finland, 2022, pp. 1-5.
	
	\bibitem{gao2022TIT}
	J. Gao, Y. Wu, S. Shao, W. Yang, and H. V. Poor, ``Energy efficiency of massive random access in MIMO quasi-static Rayleigh fading channels with finite blocklength," \emph{IEEE Trans. Inf. Theory}, vol. 69, no. 3, pp. 1618-1657, March 2023.
	
	\bibitem{yury2019ACSSC}	
	 S. S. Kowshik, K. Andreev, A. Frolov, and Y. Polyanskiy,  ``Short-packet low-power coded access for massive MAC,” in \emph{Proc. Asilomar Conf. Signals, Syst., Comput.}, Pacific Grove, CA, USA, Nov. 2019, pp. 827–832.
	 
	 \bibitem{Amal2019ICASSP}
	 V. K. Amalladinne, K. R. Narayanan, J.-F. Chamberland, and D. Guo, ``Asynchronous neighbor discovery using coupled compressive sensing,” in \emph{Proc. IEEE Int. Conf. Acoust., Speech Signal Process. (ICASSP)}, Brighton, UK, 2019, pp. 4569–4573.
	 
	 \bibitem{Decu2022GC}
	 A. Decurninge, P. Ferrand and M. Guillaud, ``Massive random access with tensor-based modulation in the presence of timing offsets," in \emph{Proc. IEEE Global Commun. Conf. (GLOBECOM)}, Rio de Janeiro, Brazil, 2022, pp. 1061-1066.
	 
	 \bibitem{sun2022TWC}
	 G. Sun, Y. Li, X. Yi, W. Wang, X. Gao, L. Wang, F. Wei, and Y. Chen, ``Massive grant-free OFDMA with timing and frequency offsets,” \emph{IEEE Trans. Wireless Commun.}, vol. 21, no. 5, pp. 3365– 3380, May 2022.
	 
	 \bibitem{yuan2021VTC}
	 Z. Yuan, Z. Li, W. Li, Y. Ma, and C. Liang, ``Contention-based grant-free transmission with extremely sparse orthogonal pilot scheme,"  in \emph{Proc. IEEE Veh. Technol. Conf. (VTC-Fall)}, Norman, OK, USA, 2021, pp. 1-6.
	 
	 \bibitem{Ahma2022isit}
	 M. J. Ahmadi and T. M. Duman, ``Unsourced random access with a massive MIMO receiver using multiple stages of orthogonal pilots," in \emph{Proc. IEEE Int. Symp. Inform. Theory (ISIT)}, Espoo, Finland, 2022, pp. 2880-2885.
	 
	  \bibitem{ping2006twc}
	 L. Ping, L. Liu, K. Wu, and W. K. Leung, ``Interleave division multiple access,” \emph{IEEE Trans. Wireless Commun.}, vol. 5, no. 4, pp. 938–947, Apr. 2006. 
	 
	 
	 
	
\end{thebibliography}
\end{document}